\documentclass[prl,preprintnumbers,nofootinbib,floatfix,twocolumn,notitlepage]{revtex4-1}

\usepackage{textcase}
\usepackage{graphicx,epsfig,psfrag,amssymb,hyperref}
\usepackage{multirow}
\usepackage{color,graphicx,epsfig,psfrag,amsmath,empheq}
\usepackage{bm}
\usepackage{mathrsfs,amsfonts,, color}
\usepackage[caption=false]{subfig}
\usepackage{hepunits}

\def\pt         {\mbox{$p_{\rm T}$}\xspace}
\newcommand{\tev}{\ensuremath{\mathrm{\: Te\kern -0.1em V}}\xspace}
\newcommand{\gev}{\ensuremath{\mathrm{\: Ge\kern -0.1em V}}\xspace}
\newcommand{\mev}{\ensuremath{\mathrm{\: Me\kern -0.1em V}}\xspace}

\def\invfb   {\ensuremath{\mbox{\,fb}^{-1}}\xspace}
\def\pythia     {\mbox{\textsc{Pythia}}\xspace}
\def\powheg     {\mbox{\textsc{PowhegBox}}\xspace}
\def\amc     {\mbox{aMC@NLO}\xspace}

\def\Zcj {\mbox{$Z^c_j$}\xspace}

\def\Zc {\mbox{$Zc$}\xspace}
\def\Zj {\mbox{$Zj$}\xspace}

\begin{document}

\title{A direct probe of the intrinsic charm content of the proton}

\author{Tom Boettcher}
\email{tboettch@mit.edu}

\author{Philip Ilten}
\email{philten@cern.ch}

\author{Mike Williams}
\email{mwill@mit.edu}
\affiliation{Laboratory for Nuclear Science, Massachusetts Institute of Technology, Cambridge, MA 02139, U.S.A.}

\begin{abstract}

Measurement of $Z$ bosons produced in association with charm jets $(Zc)$ in proton-proton collisions in the forward region provides a direct probe of a potential non-perturbative (intrinsic) charm component in the proton wave function. 
We provide a detailed study of the potential to measure $Zc$ production at the LHCb experiment in Runs 2 and 3 of the LHC.  The sensitivity to valence-like (sea-like) intrinsic charm is predicted to be $\langle x \rangle_{\rm IC} \gtrsim 0.3\%(1\%)$.  
 The impact of intrinsic charm on Higgs production at the LHC, including $Hc$, is also discussed in detail.

\end{abstract}

\maketitle

\section{Introduction}

Whether the proton wave function contains an intrinsic charm (IC) component is a topic of considerable interest (see Ref.~\cite{Brodsky:2015fna} for a review).  
In the absence of IC, the charm ($c$) parton distribution function (PDF) arises entirely due to perturbative gluon radiation; however, a $|uudc\bar{c}\rangle$ component to the proton wave function is also possible.    
There is substantial theoretical interest in the role that non-perturbative dynamics play in the nucleon sea~\cite{Franz:2000ee,Broadsky:2012rw,Gong:2013vja}. 
Furthermore, the presence of IC in the proton would affect the cross sections of many processes at the LHC either directly, from $c$ or $\bar{c}$ initiated production; or indirectly, since altering the $c$ PDF would affect other PDFs via the momentum sum rule.  
For example, Higgs boson production could be affected by a few percent, largely due to changes in the gluon PDF.  
The cross sections relevant for direct dark matter detection are sensitive to IC if the interaction is mediated by the Higgs boson~\cite{Freeman:2012ry}.  
 IC would also affect both the rate and kinematical properties of $c$-hadrons produced by cosmic-ray proton interactions in the atmosphere.  Semileptonic decays of such $c$-hadrons provide an important background to astrophysical neutrinos~\cite{Aartsen:2013jdh,Gauld:2015kvh}.  

A number of studies have been performed to determine if -- and at what level -- IC exists in the proton.  
Measurements of $c$-hadron production from deep inelastic scattering (DIS)~\cite{Aubert:1982tt}, where the typical momentum transfer is $Q \approx 1-10\gev$, have been interpreted as evidence for percent-level $c$-content in the proton at large momentum fraction ($x$)~\cite{Hoffmann:1983ah,Harris:1995jx,Steffens:1999hx}. 
If the $c$ PDF is entirely perturbative in nature, much smaller $c$ content at large $x$ is expected; whereas, valence-like charm content in the proton could explain the DIS results.   
However, global PDF analyses tend to either provide inconclusive results on IC~\cite{Dulat:2013hea}, or claim that IC is excluded at a level significantly less than 1\%~\cite{Jimenez-Delgado:2014zga}.  
There is tension between some data sets applicable to such analyses where they overlap kinematically.  This has led to global PDF fitters choosing either which data sets to consider, or how to handle the inherent tension between data sets in their studies.  
Low-energy fixed-target experiments are in principle sensitive to large-$x$ IC, but inclusion of such low-$Q$ data requires careful treatment of hadronic and nuclear effects. Therefore, many authors have chosen to exclude these data.   
Such choices inevitably affect the conclusions drawn about IC. 
To date, a consensus has not been reached on whether IC exists at the percent level~\cite{Brodsky:2015uwa,Jimenez-Delgado:2015tma}.

\begin{figure}[b]
\centering
\includegraphics[width=0.49\columnwidth]{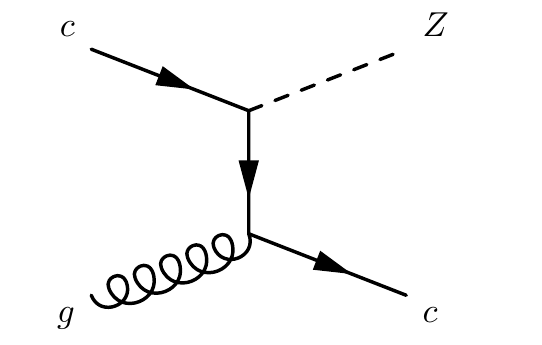}
\includegraphics[width=0.49\columnwidth]{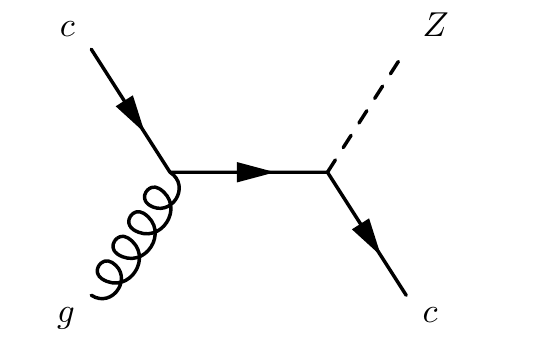}
\caption{Leading-order Feynman diagrams for $gc\!\to Zc$.}
\label{fig:diagrams}
\end{figure}

\begin{figure}[]
\includegraphics[width=0.99\columnwidth]{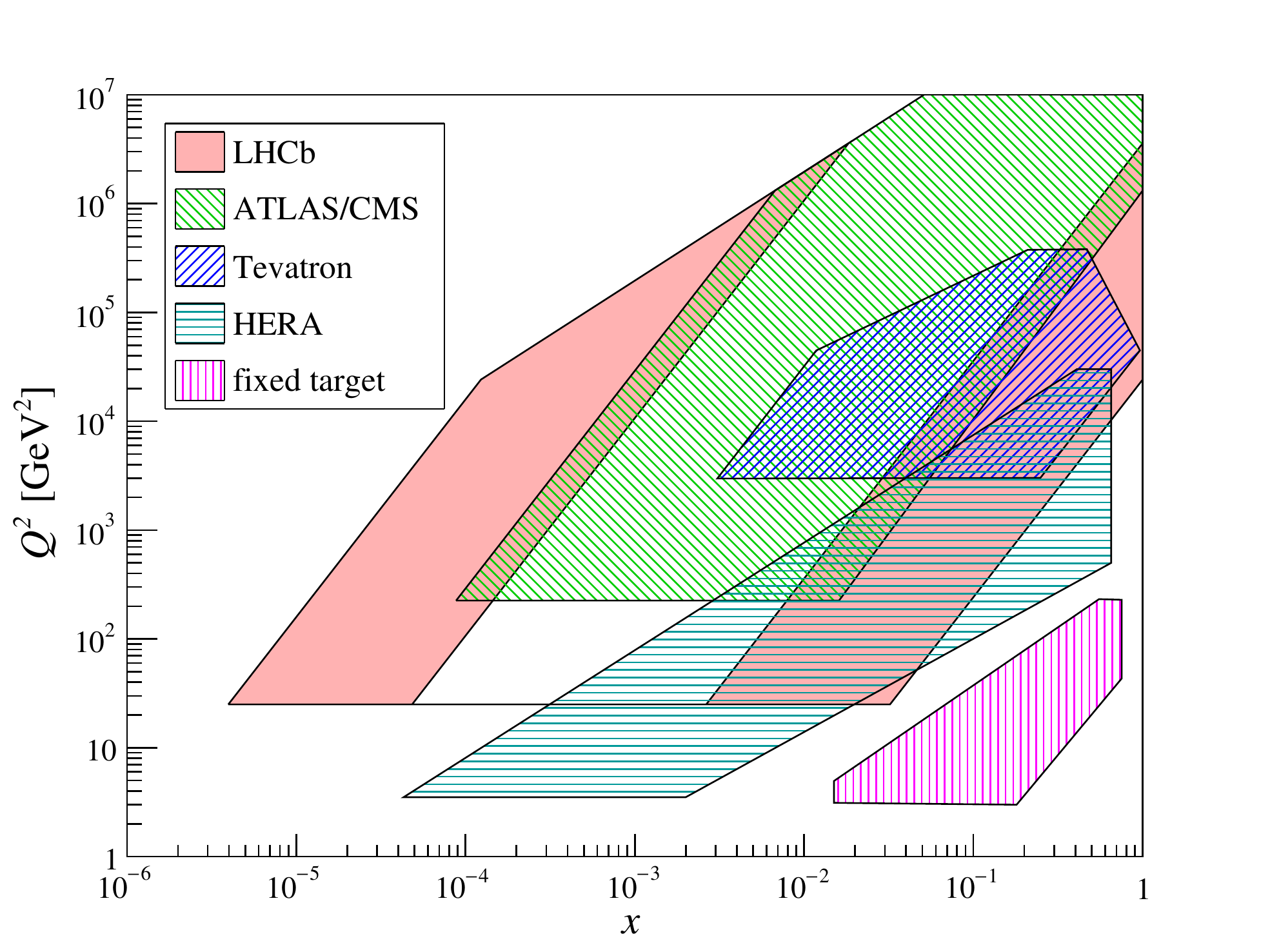}
\caption{Regions covered in $(Q^2,x)$ of various experiments. }
\label{fig:pdfs}
\end{figure}

The ideal probe of IC is a high-precision measurement of an observable with direct sensitivity to the large-$x$ charm PDF, where $Q$ is large enough such that hadronic and nuclear effects are negligible.   
Measurement of the fraction of $Z\!+$jet events where the jet originates from a $c$ quark, $\Zcj\equiv\sigma(\Zc)/\sigma(\Zj)$, in the forward region at the LHC can provide such a probe.  
Production of \Zc may proceed via $gc\!\to Zc$ (see Fig.~\ref{fig:diagrams}); is inherently at large $Q$ satisfying the constraints of Ref.~\cite{Blumlein:2015qcn} (due to the large $Z$ mass); and   
at forward rapidities requires one initial parton to have large $x$, while the other must have small $x$ (see Fig.~\ref{fig:pdfs}).  
Differential measurement of \Zcj provides direct sensitivity to the process $gc\!\to Zc$ for large-$x$ $c$.  
The ratio \Zcj is chosen because it is less sensitive to experimental and theoretical uncertainties than $\sigma(\Zc)$.  

In this Letter, we propose a differential measurement 
of \Zc production in proton-proton ($pp$) collisions in the forward region.   We show that using data that will be collected in Runs~2 and 3 of the LHC, the LHCb experiment will be highly sensitive to both valence-like and sea-like IC. 
While measurement of $\sigma(Zc)$ in the central region has previously been proposed to study IC~\cite{Dulat:2013hea}, we will show that the impact of IC is larger in the forward region and that the LHCb detector is best suited to making a precise measurement of $\sigma(Zc)$.   
Finally, even in the absence of discovery of IC content in the proton, this measurement will provide a useful test of DGLAP evolution for $c$  quarks from low-$Q$ DIS measurements up to the electroweak scale.

\section{$\mathbf{\Zc}$ Production} 

We calculate \Zcj at next-to-leading order (NLO) using the so-called VFNS CT14 next-to-NLO (NNLO) PDF set~\cite{CT14} and the \Zj \powheg matrix element~\cite{Alioli:2010qp}, 
and cross-check our results with \amc~\cite{Alwall:2014hca}; 
showering is performed via \pythia~\cite{Sjostrand:2014zea} using the {\sc Powheg}~\cite{Nason:2004rx} and {\sc FxFx}~\cite{Frederix:2012ps} methods for our baseline and cross-check calculations, respectively. 
Hadronization is also performed with \pythia, while hadrons are decayed via {\sc EvtGen}~\cite{Lange:2001uf} interfaced to {\sc Photos}~\cite{Golonka:2005pn}. 
We only consider the decay $Z\!\to\mu\mu$, and in all cases $Z$ denotes $Z/\gamma^*$ where $60 < m(\mu\mu) < 120\gev$.  

The leading-order contribution to \Zc production is $gc\!\to Zc$ as shown in Fig.~\ref{fig:diagrams}; however, at NLO there are also sizable contributions from $gc\!\to Zcg$, $gg\!\to Zc\bar{c}$, $qc\!\to Zcq$, and $q\bar{q}\!\to Zc\bar{c}$. 
The theory uncertainty on \Zcj is a combination of PDF, factorization and renormalization scale, and strong-coupling ($\alpha_s$) uncertainties, where the PDF contribution is found to be dominant (since the others largely cancel in the ratio).   
Charm jets are identified at the particle level by the presence of a long-lived $c$-hadron with transverse momentum $\pt > 2\gev$ produced promptly in the $pp$ collision.  

\begin{figure}[]
\includegraphics[width=0.99\columnwidth]{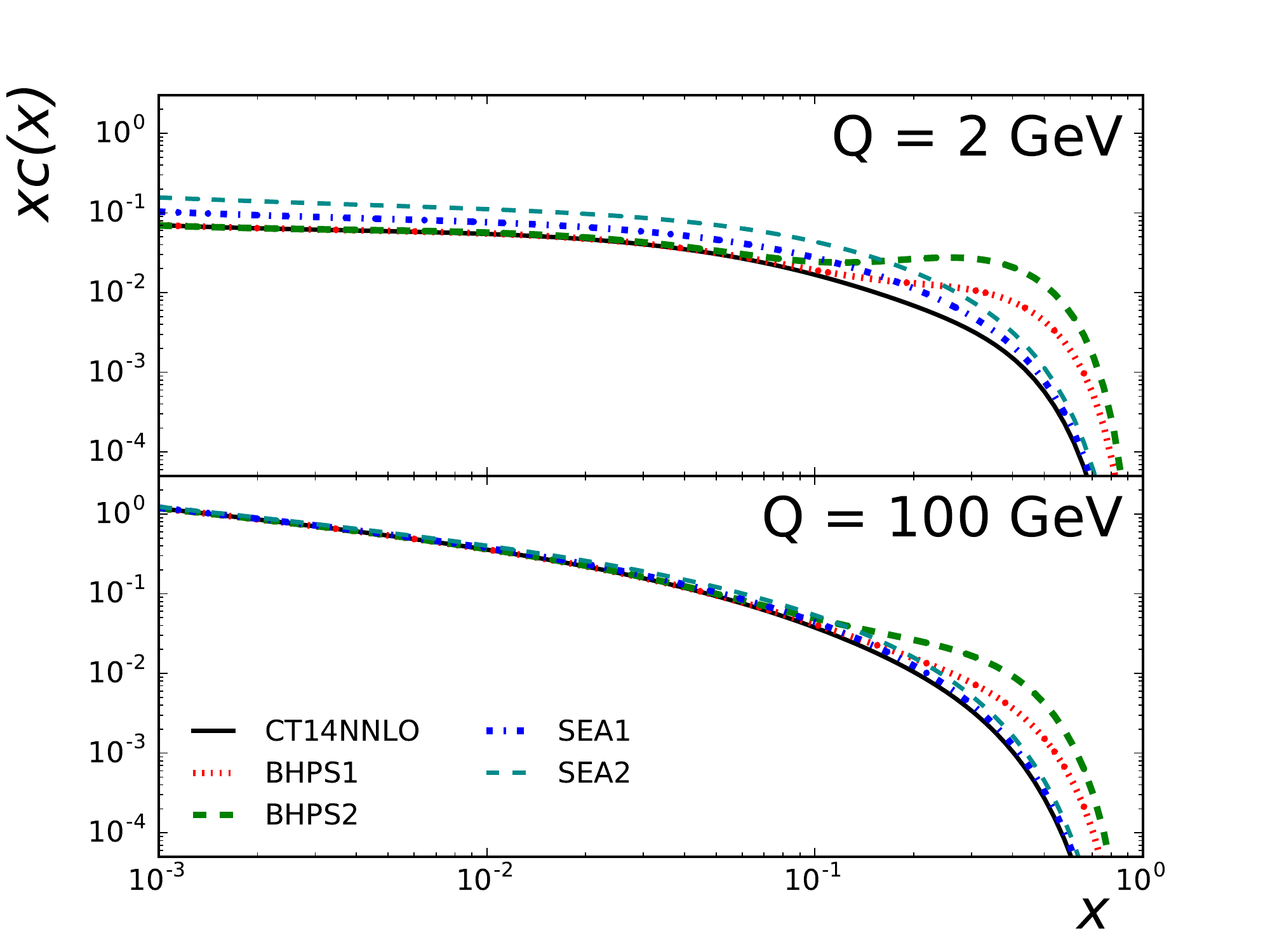}
\caption{IC PDFs considered~\cite{Dulat:2013hea} shown at low and high $Q$.}
\label{fig:icmodels}
\end{figure}

The CT14 global analysis turns on the $c$ and $\bar{c}$ PDFs at $Q=m(c)$, {\em i.e.}\ at the charm mass, with initial distributions  $c(x,m(c))=\bar{c}(x,m(c))$ consistent with NNLO matching. 
At NLO, $c(x,m(c))=\bar{c}(x,m(c))=0$, while at NNLO they are of $\mathcal{O}(\alpha_s^2)$.   
Additional $c$ content is generated by gluonic radiation for $Q>m(c)$.  
Following Ref.~\cite{Dulat:2013hea}, we consider two categories of non-perturbative IC models: 
(BHPS) valence-like, inspired by the light-cone picture of nucleon structure~\cite{Brodsky:1980pb,Brodsky:1981se}; 
and (SEA) sea-like, where IC $\propto \left[\bar{u}(x,Q_0)+\bar{d}(x,Q_0)\right]$ at an initial scale $Q_0 < m(c)$.  
In each model, the IC content is considered in addition to the perturbative charm contribution.  
For each IC category,  two values of the mean momentum fraction of the IC PDF at $Q=m(c)$,  $\langle x \rangle_{\rm IC} \equiv \int_0^1 x\, {\rm IC}(x,m(c)){\rm d}x$, are considered: roughly the maximum $\langle x \rangle_{\rm IC}$ value that is consistent with the global PDF analysis of CT14~\cite{CT14}, and a smaller IC contribution
(see Tab.~\ref{tab:icmodels} and Fig.~\ref{fig:icmodels}).   
Many other IC models exist (see, {\em e.g.}, Ref.~\cite{Hobbs:2013bia}); however, we only consider the BHPS and SEA models as this is sufficient to demonstrate the impact of both low-$x$ and high-$x$ IC.  

\begin{table}
\begin{tabular}{cc|cc}
\hline \hline
valence-like model & $\langle x \rangle_{\rm IC}$ & sea-like model & $\langle x \rangle_{\rm IC}$ \\
\hline
BHPS1 & 0.6\% & SEA1 & 0.6\% \\
BHPS2 & 2.0\% & SEA2 & 1.5\%\\
\hline \hline
\end{tabular}
\caption{IC models considered~\cite{Dulat:2013hea}.}
\label{tab:icmodels}
\end{table}

\begin{figure*}[]
\includegraphics[width=0.99\columnwidth]{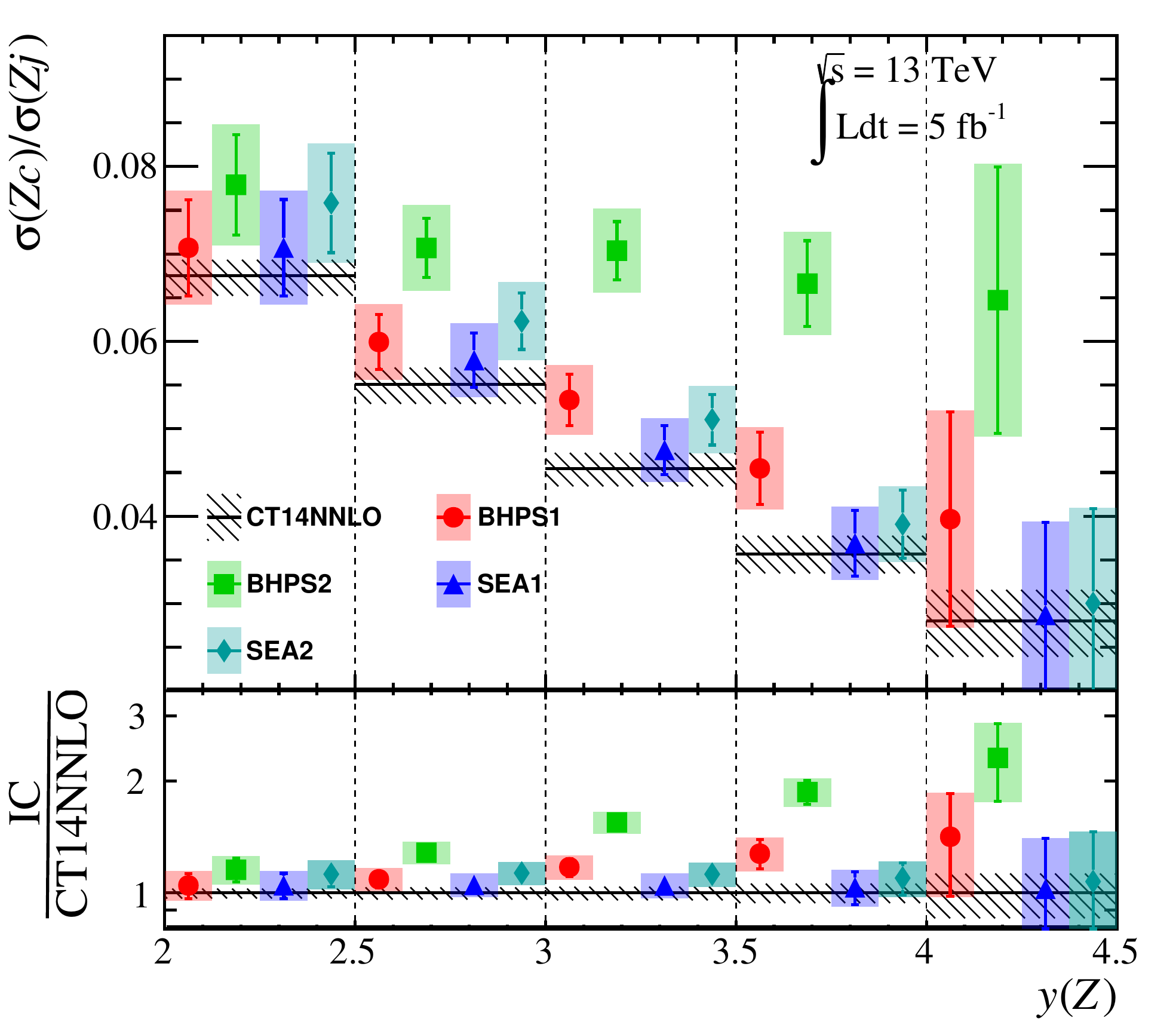}
\includegraphics[width=0.99\columnwidth]{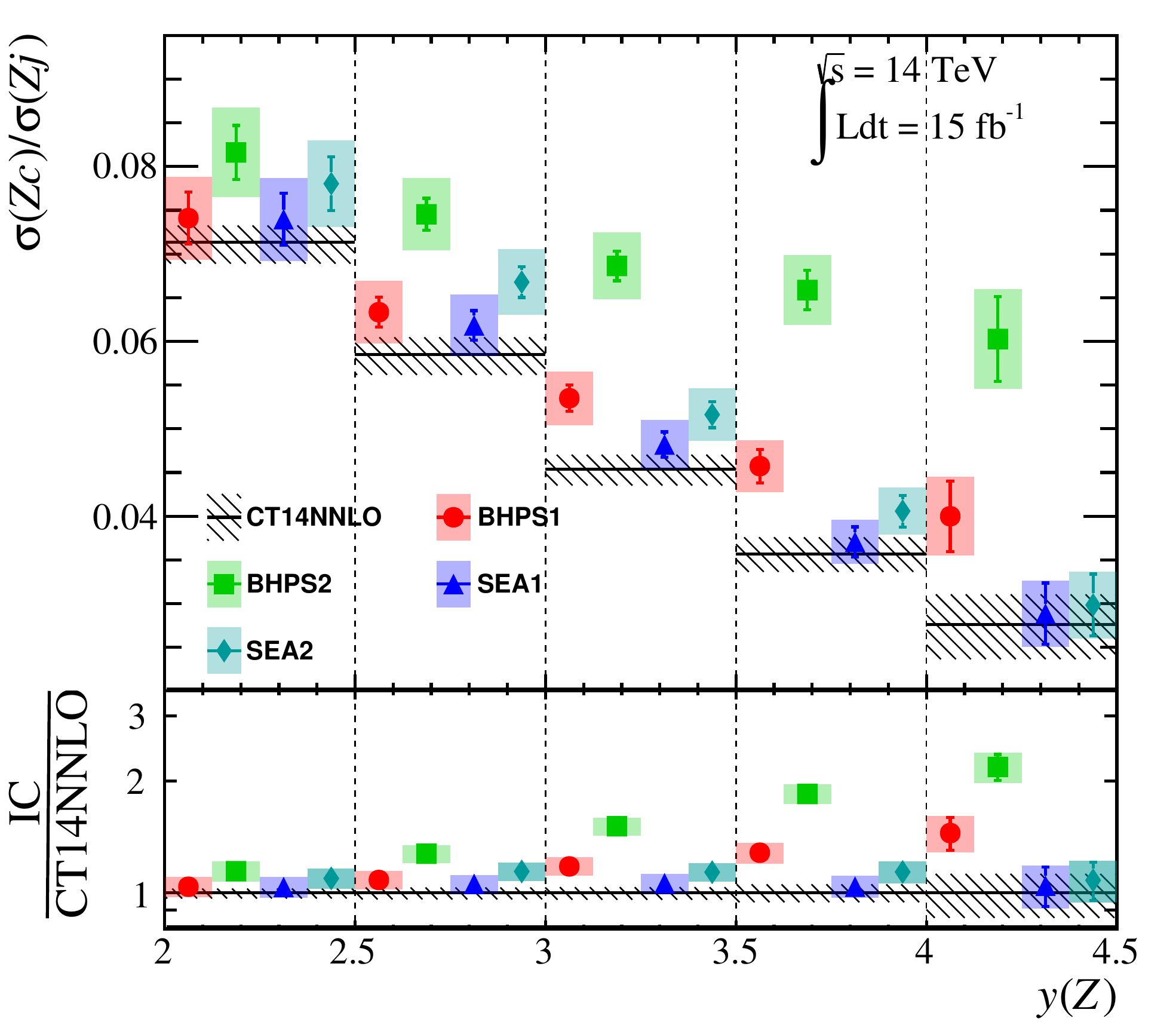}
\caption{
Predictions for \Zcj for (left) Run~2 and (right) Run~3. For each IC model prediction, the expected experimental statistical uncertainty is shown by the error bar, while the expected total experimental uncertainty is given by the shaded box. 
The total theory uncertainty is shown as the hashed box around the CT14NNLO-based prediction.  
The bottom plots show the relative impact of the various IC PDFs; note the log scale.   
{\em N.b.}, the experimental systematic uncertainty is nearly 100\% correlated across $y(Z)$ bins; the IC-model predictions are staggered within in each bin to aid readability.  
}
\label{fig:zcj}
\end{figure*}

\section{Selection and Detector Performance}

The LHCb detector is a single-arm spectrometer covering the forward region of $2<\eta <5$~\cite{Alves:2008zz,Aaij:2014jba}.
The detector, built to study the decays of hadrons containing $b$ and $c$ quarks, includes a high-precision charged-particle tracking system.  
The silicon-strip vertex locator (VELO) that surrounds the $pp$ interaction region measures heavy-flavor hadron lifetimes with an uncertainty of about 50\,fs~\cite{LHCbVELOGroup:2014uea}.
Different types of particles are distinguished using information from two ring-imaging Cherenkov (RICH) detectors, an electromagnetic and hadronic calorimeter system, and a system of muon chambers~\cite{LHCb-DP-2012-002}. 

Our analysis assumes that the LHCb detector performance will be equivalent in Runs~1 and 2 of the LHC, and that LHCb will collect 5\invfb of data at $\sqrt{s}=13\tev$ in Run~2.  
For Run~3, we take the detector performance from the LHCb subsystem technical design reports~\cite{LHCb-TDR-013,LHCb-TDR-014,LHCb-TDR-015,LHCb-TDR-016}, and assume 15\invfb is collected at $\sqrt{s}=14\tev$.    

LHCb has demonstrated the ability to make precise measurements of $Z$ boson production~\cite{Aaij:2015gna,Aaij:2015zlq,LHCb-PAPER-2014-055}.  Here, we assume only the decay $Z\!\to\mu\mu$ is used as it provides the most precise experimental measurements.  Following Ref.~\cite{Aaij:2015gna}, we define the muon fiducial region to be $\pt(\mu) > 20\gev$ and $2 < \eta(\mu) < 4.5$ for Run~2.
For Run~3, the $\eta(\mu)$ region is extended to $2 < \eta(\mu) < 5$ due to the improved tracking coverage upstream of the magnet that will be provided by the so-called UT system~\cite{LHCb-TDR-015}.   
$Z$ boson candidates are required to satisfy $60 < m(\mu\mu) < 120\gev$.  
Furthermore, we assume that quality criteria are imposed on the track and muon, and take the efficiency of such requirements from Ref.~\cite{Aaij:2015gna}.  

Jets are clustered using the anti-$k_{\rm T}$ algorithm~\cite{1126-6708-2008-04-063} with $R=0.5$ as implemented in {\sc FastJet}~\cite{fastjet}.  Only visible final-state particles within LHCb acceptance are clustered.  
As in Refs.~\cite{Aaij:2015yqa,Aaij:2015cha,Aaij:2015mwa}, jets are required to satisfy $\pt(j) > 20\gev$ and $2.2 < \eta(j) < 4.2$ to ensure nearly uniform jet reconstruction and $c$-jet-identification efficiencies, and only the highest-\pt jet in each event is considered (all other jets are ignored).  
  Ref.~\cite{Aaij:2015cha} demonstrates that migration of events in and out of this fiducial region due to detector response has negligible impact on the production ratios studied here; therefore, jet \pt resolution effects are not considered in this study.  
LHCb applies criteria to remove fake jets with a 96\% efficiency~\cite{LHCb-PAPER-2013-058}; we assume these will also be applied in Runs~2 and 3.  
LHCb discards very high-occupancy events as part of its online data-taking optimization.  We again assume that this effect will be the same in the future as it was in Run~1, and reduce the expected signal yields by 10\%~\cite{LHCb-PAPER-2013-058}.

A key aspect of the proposed measurement in this Letter is the ability to efficiently identify (or tag) $c$-jets.  LHCb has demonstrated the ability to identify heavy-flavor-hadron decay vertices in jets with a $\approx 0.3\%$ fake rate~\cite{Aaij:2015yqa}.  
Furthermore, LHCb can determine the $c$-jet and $b$-jet yields each with percent-level precision.  
While we expect the $c$-jet identification efficiency to improve in future LHCb data taking, here we assume that it is $\epsilon_{\rm tag}(c)\approx 25\%$ as it was in Run~1~\cite{Aaij:2015yqa}.

The values of  $\epsilon_{\rm tag}(c)$ and $\epsilon_{\rm tag}(b)$ were measured simultaneously by LHCb  using heavy-flavor-jet enriched data samples.  
No assumptions were made about the efficiency values in data, {\em c.f.}\ simulation, which led to a high degree of anti-correlation between 
the $\epsilon_{\rm tag}(c)$ and $\epsilon_{\rm tag}(b)$ measurements; each was assigned a 10\% relative uncertainty.   
Ref.~\cite{Gauld:2015qha} shows that the ratio $\sigma(c\bar{c})/\sigma(b\bar{b})$ is robust with respect to higher-order QCD corrections.  Therefore, the ratio   $\epsilon_{\rm tag}(c)/\epsilon_{\rm tag}(b)$  can be precisely measured in a data-driven way in an analysis similar to Ref.~\cite{LHCb-PAPER-2014-023}, removing the large anti-correlation effect. 
In this study, we assume that a 5\% relative uncertainty is achieved on $\epsilon_{\rm tag}(c)$ in Runs 2 and 3.   
Finally, background to \Zj events will be at the sub-percent level~\cite{LHCb-PAPER-2013-058} and approximately cancels in the ratios studied here, so is ignored.

\section{Expected Sensitivity}

Figure~\ref{fig:zcj} shows the expected distributions and precision on \Zcj versus the rapidity ($y$) of the $Z$ for each IC model considered compared to the no-IC prediction.  The expected results from the LHCb experiment after Runs~2 and 3 of the LHC are each shown assuming the detector performs as described above.  Most experimental and theoretical uncertainties approximately cancel in this ratio.
The dominant contribution to the experimental systematic uncertainty comes from how well $\epsilon_{\rm tag}(c)$ can be measured in data. 
There will also likely be $\approx 1\%$ contributions from various ratios of effects, {\em e.g.}, the efficiency of event-occupancy requirements in \Zc compared to \Zj events.  These can each be studied using data-driven methods and are not expected to increase the total systematic uncertainty significantly.

From Fig.~\ref{fig:zcj} one can see that valence-like IC has a dramatic impact on \Zcj at large $y(Z)$, while sea-like IC mostly affects \Zcj at small $y(Z)$.  
Both the shape and size of the measured \Zcj versus $y(Z)$ distribution can be used to study IC($x$). 
By the end of Run~3, we estimate that LHCb will be sensitive to IC of the type found in BHPS models for $\langle x \rangle_{\rm IC} \gtrsim 0.3\%$, and to that found in SEA models for $\langle x \rangle_{\rm IC} \gtrsim 1\%$.
The impact of valence-like IC on \Zcj in the forward region is so large that discovery of IC will be possible already in Run~2 for $\langle x \rangle_{\rm IC} \gtrsim 1\%$.
If such a valence-like IC component is observed, then it may even be possible in Run~3 to investigate the $c$ and $\bar{c}$ PDFs separately by tagging the charge of the $c$-jet.  
Predictions of \Zcj in the central region (probed by ATLAS and CMS) are provided as Supplemental Material to this Letter~\cite{supp}. 
As expected, the impact of valence-like IC is greatly reduced, while sea-like IC affects \Zcj in a similar way as in forward region.

We conclude our discussion on measuring IC by considering the ratio $\sigma\left(\gamma c\right)/\sigma\left(\gamma j\right)$, {\em i.e.}\ replacing the $Z$ boson with a final-state photon, which would permit probing lower values of $Q$~\cite{Stavreva:2009vi,Bednyakov:2013zta,Rostami:2015iva}.  
Such measurements have been made at the Tevatron~\cite{D0:2012gw,Aaltonen:2013ama} and are suggestive of IC~\cite{Mesropian:2014kfa}. 
The LHCb calorimeter system is not well suited to studying high-energy photons; however, LHCb has demonstrated that it can reconstruct and precisely measure the properties of $\gamma$ conversions to $e^+e^-$~\cite{Aaij:2013dja}.  It may be possible to measure $\gamma c$ production using converted photons at LHCb~\cite{Ward}.  We encourage studying this possibility.    

For the large valence-like IC scenario, a sizable intrinsic beauty (IB) component may also be present.  While IB is expected to be suppressed relative to IC by a factor of roughly $[m(c)/m(b)]^2$~\cite{Lyonnet:2015dca}, it may still be possible to observe a large valence-like IB component by studying $Zb$ production.  Given that the jet-tagging algorithm developed by LHCb simultaneously determines both the $b$-jet and $c$-jet yields, we expect that both \Zcj and $Z^b_j$ will be measured in the same analysis with about the same precision.

\section{Impact on Other Processes}

The IC content of the proton directly affects the production cross sections of many processes at the LHC.  
For example, valence-like IC content increases $W$ boson production due to an increased probability for $cs\!\to W$ scattering~\cite{Halzen:2013bqa}.   
Similarly, the rate at which the hypothetical charged Higgs boson is produced in $pp$ collisions is highly sensitive to IC~\cite{Lai:2007dq}.
Furthermore, an increase in the charm component of the proton must be balanced by a decrease in the other components.
This results in IC indirectly affecting many production cross sections via the momentum sum rule.  

\begin{figure}[]
\includegraphics[width=0.9\columnwidth]{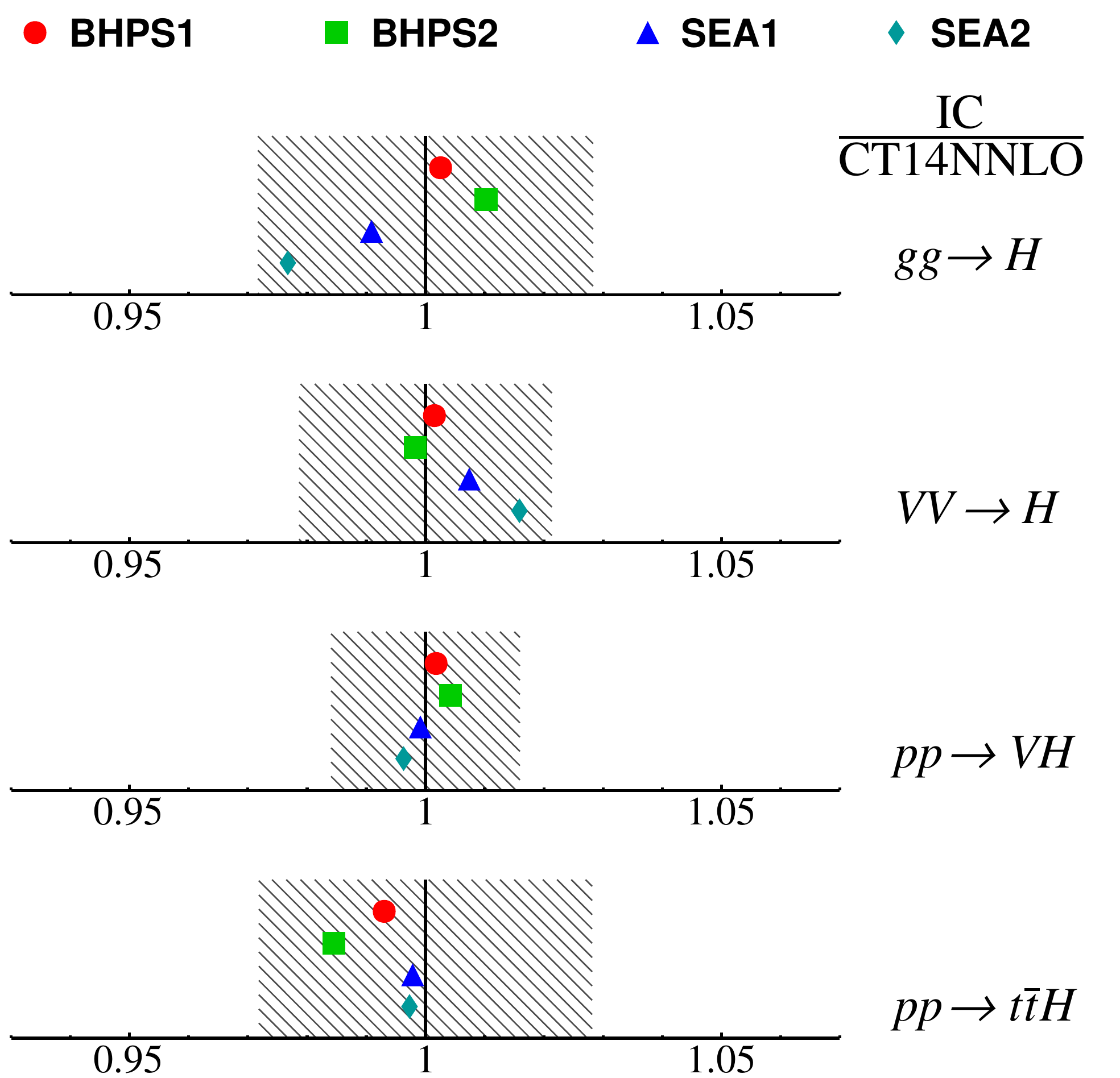}
\caption{
Impact of IC models on Higgs production in the central region.  The hashed boxes show the PDF uncertainties. 
}
\label{fig:higgs}
\end{figure}

Figure~\ref{fig:higgs} shows the impact on the major Higgs production cross sections within the acceptance of the ATLAS and CMS detectors, assuming a SM Higgs boson (details on these calculations are provided in Ref.~\cite{supp}). 
Since the PDF uncertainties do not include a contribution due to the assumption of no IC in the proton, one should view the shift due to IC as an additional uncertainty in each Higgs-production process.  
Higgs production in the central region via gluon fusion ($gg\!\to H$) is affected by $\lesssim 1\%$ by valence-like IC, but by up to $\approx 2.5\%$ by sea-like IC.  
Higgs production via vector boson fusion ($VV\!\to H$) is also affected by sea-like IC by $\approx 2\%$ but in the opposite way.  
This is expected since adding sea-like IC increases the quark content of the proton while decreasing its gluon content.  
Associated production of $t\bar{t}H$ is affected by up to 1.5\% by valence-like IC due to the large $Q$, hence large $x$ of one parton, of this $gg$-initiated process.  
The predicted sensitivity to IC at LHCb in Run~3 will be sufficient to constrain the effect of valence-like (sea-like) IC on all major Higgs production processes to be $\lesssim 0.5\% (1\%)$. 
We also note that the presence of IC in the proton could result in diffractive Higgs production~\cite{Brodsky:2006wb,Brodsky:2007yz}. 

Ref.~\cite{Brivio:2015fxa} suggests measuring $Hc$ production as a way of probing the charm Yukawa coupling $(Y_c)$.  
It is worth noting that the impact of IC on $\sigma(Hc)$ is comparable to that of a SM-like value of $Y_c$.  
For both the BHPS2 and SEA2 IC models, $\sigma(Hc)$ with $Y_c = 0$ is about the same as for the no-IC scenario with $Y_c \approx 0.7\, Y_c^{\rm SM}$.  
Similarly, if $Y_c = Y_c^{\rm SM}$ then these IC models would increase $\sigma(Hc)$ by as much as a $\approx 25\%$ increase in $Y_c$. 
Therefore, placing constraints on IC will be a vital component of determining $Y_c$ using $Hc$ production. 

\section{Summary}

In summary, 
measurement of $\sigma(\Zc)/\sigma(\Zj)$ in $pp$ collisions in the forward region provides a direct probe of a potential intrinsic charm component in the proton wave function. 
We predict that using data collected in Runs~2 and 3 of the LHC, the LHCb experiment will be sensitive to valence-like IC with 
$\langle x \rangle_{\rm IC} \gtrsim 0.3\%$, and sea-like IC for $\langle x \rangle_{\rm IC} \gtrsim 1\%$.
This sensitivity is sufficient to discover, in the context of a global PDF analysis, the IC predicted by light-cone calculations~\cite{Brodsky:1980pb,Brodsky:1981se}, and to constrain the uncertainty on the affect of IC on Higgs boson production at the LHC to $\lesssim 1\%$. 
We reiterate that even in the absence of IC, this measurement will provide a useful test of DGLAP evolution for $c$ quarks from low-$Q$ DIS measurements up to the electroweak scale.  
Finally, a similar analysis of $\sigma(Wc)/\sigma(Wj)$ versus $\eta(\mu)$ can be performed at LHCb to probe the large-$x$ strange PDFs, where the charge of the $W$ boson determines whether the initial parton was an $s$ or $\bar{s}$.  Given the large ratio of $\sigma(Wc)/\sigma(Zc)$, precision measurement of the charge asymmetry in $Wc$ production should be possible already in Run~2.

\section*{Acknowledgements}
\noindent
We thank S.J.~Brodsky, R.~Gauld, P.~Koppenburg, W.~Melnitchouk, and F.~Olness for helpful advice and feedback. 
P.I.\ and M.W.\ are supported by the U.S.\ National Science Foundation grant PHY-1306550.

\section*{Supplemental Material}

\begin{figure}[]
\includegraphics[width=0.99\columnwidth]{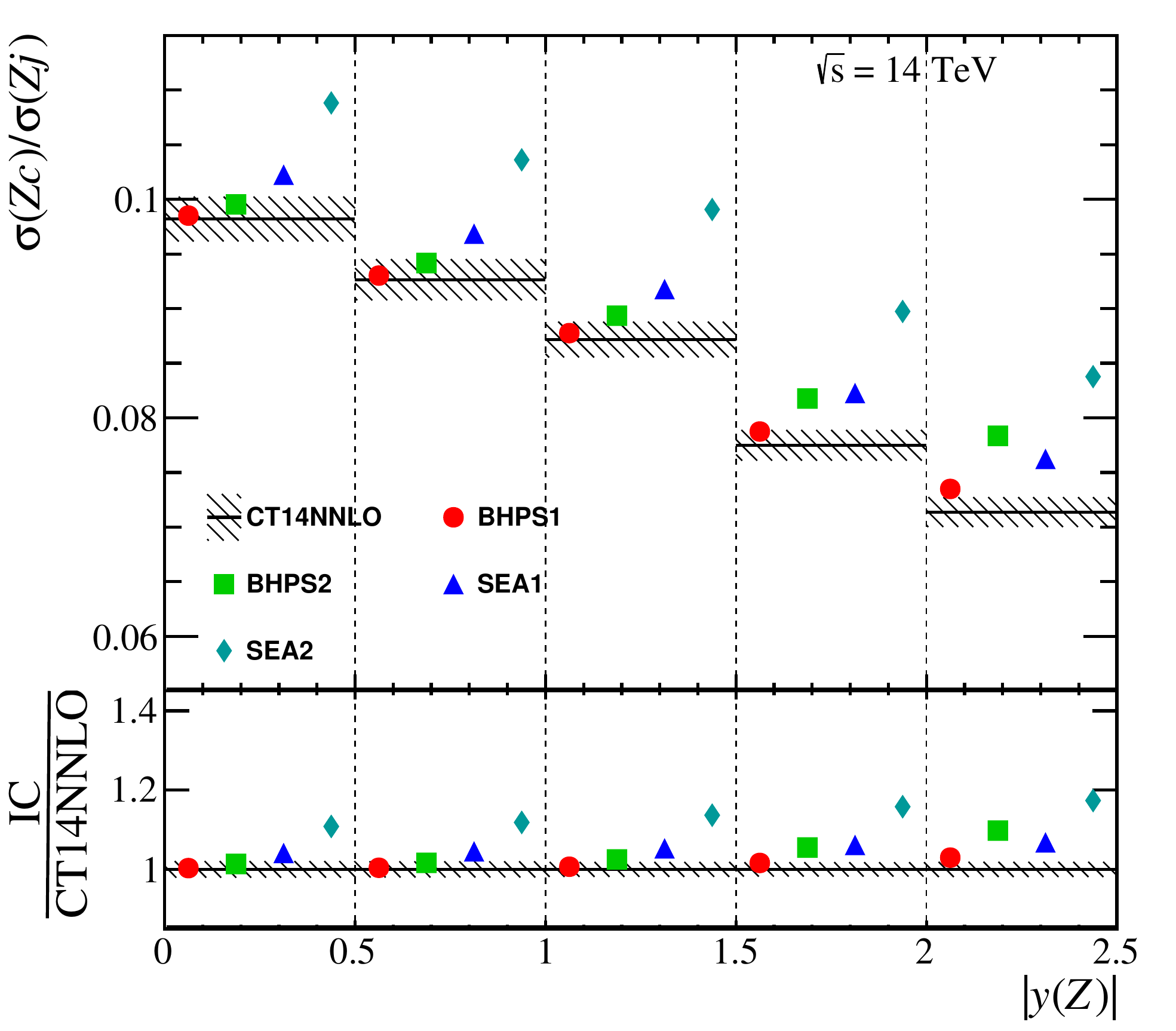}
\caption{
Predictions of \Zcj for the central region.  No experimental error bars are provided, as these will be driven by the systematic uncertainty on determining the tagged $c$-jet yield and $c$-tagging efficiency.  Only the SEA2 model affects \Zcj by $\gtrsim 10\%$. {\em N.b.}, due to the small effect of IC in the central region, the lower panel here is presented using a linear scale {\em c.f.}\ the log scale used in Fig.~\ref{fig:zcj}.   
}
\label{fig:zcj_central}
\end{figure}

Figure~\ref{fig:zcj_central} shows the \Zcj predictions for the central region.  These calculations are performed in the same manner as those for within the LHCb acceptance, but with the muon and jet $\eta$ requirements changed to $|\eta| < 2.5$.  Given the large luminosity that ATLAS and CMS expect to collect, the experimental uncertainty on \Zcj will be driven by the $c$-jet tagging.  Only the SEA2 model affects \Zcj by $\gtrsim 10\%$.   
With enough luminosity -- and assuming a $c$-tagging efficiency of about $10\%$ -- it may be possible to improve the sensitivity to valence-like IC in the central region by instead measuring \Zcj at large $\pt(Z)$~\cite{Dulat:2013hea}; however, we do not expect that equivalent sensitivity to that of LHCb can be achieved by CMS or ATLAS in Run~3.    
We note that an alternative approach for ATLAS/CMS is proposed in Ref.~\cite{Beauchemin:2014rya} that involves measuring the ratio $ZQ/WQ$, where $Q$ denotes all tagged $b$-jets and $c$-jets, at large jet \pt.

For predictions of SM Higgs boson production in the central region, the calculations are performed in the same manner as for \Zcj; however, only the relative impact of the IC models is provided and only PDF uncertainties are considered.  This is sufficient to demonstrate qualitatively how IC affects Higgs production.  
In all cases we require $|\eta(H)| < 2.5$, along with the following channel-specific requirements: 
$(VV\!\to H)$ for vector boson fusion, we require $\pt(j) > 20\gev$, $|\eta(j)| < 5$, and $|\Delta\eta(j)| > 3$;
$(pp\!\to VH)$ for vector boson associated production, we require all leptons from $W$ and $Z$ decays have $\pt(\ell) > 20\gev$ and $|\eta(\ell)| < 2.5$;
and $(pp\!\to t\bar{t}H)$ for top associated production, we require the leptons and $b$-jets have $|\eta| < 2.5$ and $\pt > 20\gev$. 

\vspace{-0.1in}
\bibliography{ic,lhcb}

\end{document}